\documentstyle[aps,prl,twocolumn,epsfig,floats,amssymb]{revtex}
\begin{document}
\draft  

\title{Measurements of Grain Motion in a Dense, Three-Dimensional Granular Fluid}
\author{Xiaoyu Yang, Chao Huan, and D. Candela}
\address{Physics Department, University of Massachusetts, Amherst, Massachusetts 01003}
\author{R. W. Mair and R. L. Walsworth}
\address{Harvard-Smithsonian Center for Astrophysics, Cambridge, Massachusetts 02138}
\date{November 29, 2001}
\maketitle

\begin{abstract}
	We have used an NMR technique to measure the short-time, three-dimensional displacement of grains in a system of mustard seeds vibrated vertically at 15$g$.
	The technique averages over a time interval in which the grains move ballistically, giving a direct measurement of the granular temperature profile.
	The dense, lower portion of the sample is well described by a recent hydrodynamic theory for inelastic hard spheres.
	Near the free upper surface the mean free path is longer than the particle diameter and the hydrodynamic description fails.
\end{abstract}

	Under suitable conditions, a noncohesive granular system such as dry sand can be maintained in a fluid-like state by continuous vibration of the container.
	A question of current interest is the extent to which such a system can be quantitatively described by a hydrodynamic theory, generalizing conventional nonequilibrium statistical mechanics for elastic systems such as liquids and gases\cite{jenkins8385,goldshtein95,sela98,garzo99,ramirez00,vannoije00}.
	The key distinction of granular systems is the continuous loss of kinetic energy during inelastic collisions of the grains, resulting in highly out-of-equilibrium states with strong gradients or time derivatives of macroscopic variables such as density and granular temperature.

	Theoretical and experimental\cite{clement91,warr95,wildman99,kudrolli00,rouyer00,wildman0001} studies of the ``vibrofluidized'' state usually have treated simplified conditions of uniform spherical particles with nearly elastic collisions, often in reduced dimensions.
	It is important to determine the extent to which theories developed for this idealized domain can describe more typical, technologically relevant systems.
	Such systems often have irregularly shaped grains with a distribution of sizes, and complex energy loss mechanisms only loosely modeled by the usual restitution coefficient\cite{alpha}.

	In this paper, we present an experimental study of a three-dimensional system of mustard seeds fluidized by vertical vibrations of the container.
	Pulsed field gradient nuclear magnetic resonance (PFG-NMR) combined with magnetic resonance imaging (MRI) was used to measure the distribution of positions and short-time displacements of the grains.
	In this way, we were able to measure both the density and granular temperature profiles of the system.
	To our knowledge, the only previous measurements of granular temperature profiles in \emph{three-dimensional} vibrofluidized systems were the recent positron-emission particle-tracking (PEPT) studies of Ref.~\onlinecite{wildman0001}.
	The PEPT and NMR techniques are complementary as the former traces the motion of a single tagged particle over long times, while the latter measures nearly instantaneous averages of the motion of many particles.
	The NMR method used here achieves better time resolution than previous PEPT work~\cite{wildman0001} and is used to study a much denser system.
	Diffusing-wave spectroscopy has also been used to study grain motion in three-dimensional systems\cite{menon97}.

	Two-dimensional systems have been extensively studied using video methods\cite{clement91,warr95,wildman99,kudrolli00,rouyer00}, and spin-tagged MRI has been used to measure much slower granular flows than studied here\cite{mueth00}.
	Seymour, et al. have used PFG-NMR/MRI to image the diffusion of grains in a rotating drum and infer the granular temperature\cite{seymour00}.
	In contrast to that work we study a simpler physical system for which first-principles hydrodynamic theories are currently available, and use more rapid gradient pulses to probe deep within the ballistic regime.
		In earlier work\cite{yang00} we reported MRI measurements of the density of a vibrofluidized granular system.
	The important advance reported here is the ability to measure the granular temperature profile.
	Despite the small size and high density of our system (roughly 50\% of the close-packing density), we find the density and granular temperature profiles to be surprisingly well described by a recently developed statistical theory\cite{garzo99} for inelastic hard spheres (IHS).
	Near the top of the system, which is a free boundary, the mean free path becomes long and the hydrodynamic description based on the IHS model breaks down.
	The horizontal and vertical degrees of freedom become decoupled, and the horizontal granular temperature grows with height despite ``heating'' only from below.
	We do not yet have an explanation for this phenomenon.

	The experimental setup was as follows.
	The sample cell was a vertical glass tube of inside diameter 9.00~mm, with a perforated Teflon bottom wall.
	It was sufficiently tall to prevent collisions of the grains with the top.
	The cell was supported at the center of the NMR probe and magnet (static field 0.975~T) and vibrated vertically by the arrangement described in Ref~\onlinecite{yang00}.
	The NMR protocol used a stimulated-echo PFG sequence with a single gradient pulse pair followed by frequency-encoded one-dimensional MRI\cite{callaghan91}.
	This allowed us to measure the joint distribution $p(\Delta u, v;\Delta t, t)$ of grain displacements $\Delta u$ along one axis during $\Delta t$ and grain positions $v$ along a second axis, as a function of $\Delta t$ and the time $t$ within the vibration cycle.
	By using very brief (150~$\mu$s) gradient pulses, the displacement interval $\Delta t$ could be made as short as 850~$\mu$s.
	The PFG sequence was not velocity compensated, so both collective motion (such as convection) and random motion could be observed\cite{seymour00,callaghan91}.

	A complete four-dimensional data set $p(\Delta u, v;\Delta t, t)$ is large, and in addition the physical conditions such as number and size of grains, vibration amplitude and frequency, and ambient gas pressure can be varied.
	For this study we focussed on a single sample state, consisting of 60 mustard seeds vibrated sinusoidally at 50~Hz and acceleration amplitude $(14.85 \pm 0.15)g$ where $g=9.80$~m/s$^2$.
	The ambient pressure was reduced below 200~mTorr\cite{vac}.
	The mustard seeds were roughly ellipsoidal with a mean diameter $\sigma=1.84$~mm, standard deviation of the diameter $\Delta\sigma=0.09$~mm and mean mass $m=4.98$~mg.

	We measured the joint distribution of horizontal or vertical displacements and vertical positions $p(\Delta x, z;\Delta t, t)$, $p(\Delta z, z;\Delta t, t)$ as well as the transverse mass profile $p(x)$ ($x$ is horizontal measured from the cell center and $z$ is vertical).
	To within the noise $p(x)$ fits the form $[1-(x/R)^2]^{1/2}$ for uniform transverse density in a cylinder of radius $R$, showing that there was no large variation of particle density with radial coordinate.

\begin{figure}
\includegraphics[width=0.9 \linewidth]{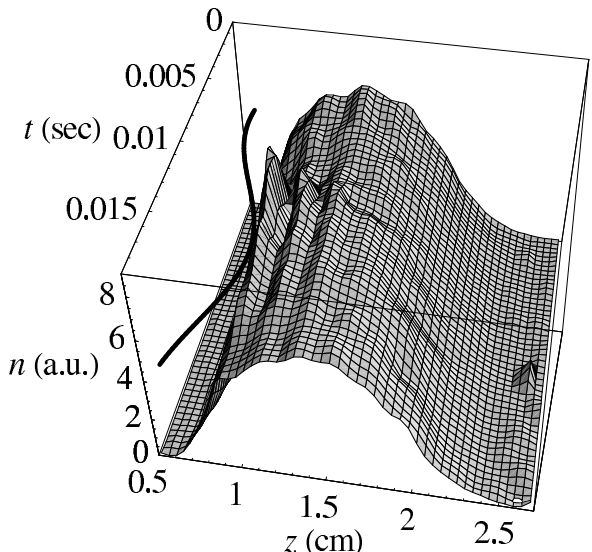}
\caption{\label{density}
	Density $n(z,t)$ of grains in arbitrary units as a function of height $z$ and time $t$ in the vibration cycle.
	The zero of the $z$ axis is arbitrary.
	The thick sinusoidal line shows the amplitude and phase of the cell motion measured by the accelerometer.
	It is drawn at the position of the cell bottom as inferred from the density data.
	For most of the vibration cycle, the sample ``floats'' well above the cell bottom.
}\end{figure}

	Figure \ref{density} shows the vertical density profile as a function of time in the vibration cycle.
	The profile is largely independent of $t$ except for a disturbance to the lower part of the sample near the peak of the vibration waveform, when the cell impacts the sample.
	These time-resolved data suggest it might be possible to describe the bulk of the sample as a steady-state statistical system ``heated'' by energy input from the vibrating cell bottom.

\begin{figure}
\includegraphics[width=0.9 \linewidth]{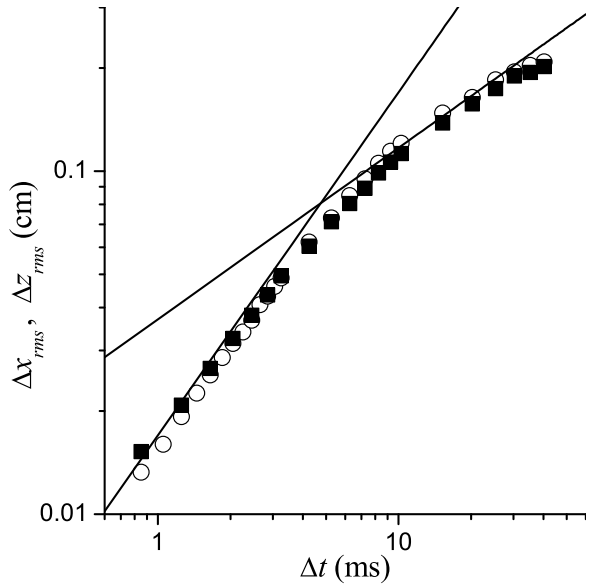}
\caption{\label{ballistic}
	Root mean square horizontal and vertical displacements $\Delta x_{rms}$ (solid squares) and $\Delta z_{rms}$ (open circles) averaged over the sample volume and time, as functions of the displacement observation interval $\Delta t$.
	The solid lines have slopes of 1.0 and 0.5, corresponding to ballistic and diffusive dependence respectively of the RMS displacement on $\Delta t$.
}\end{figure}

	The RMS grain displacements $\Delta x_{rms}, \Delta z_{rms}$ were computed from the decay of the signal magnitude vs. $q_x, q_z$, which are variables Fourier-conjugate to $\Delta x, \Delta z$ \cite{seymour00,callaghan91}.
	Figure \ref{ballistic} shows these mean-square displacements averaged over the sample volume and observation time $t$, as functions of the displacement interval $\Delta t$.
	These graphs show a crossover from ballistic ($\Delta x_{rms}, \Delta z_{rms} \propto \Delta t$) to diffusive ($\Delta x_{rms}, \Delta z_{rms} \propto \Delta t^{1/2})$) time dependence.
	The diffusive power law is not clearly established by the data, but it is the ballistic-regime information that is used for the analysis below.

\begin{figure}
\includegraphics[width=0.9 \linewidth]{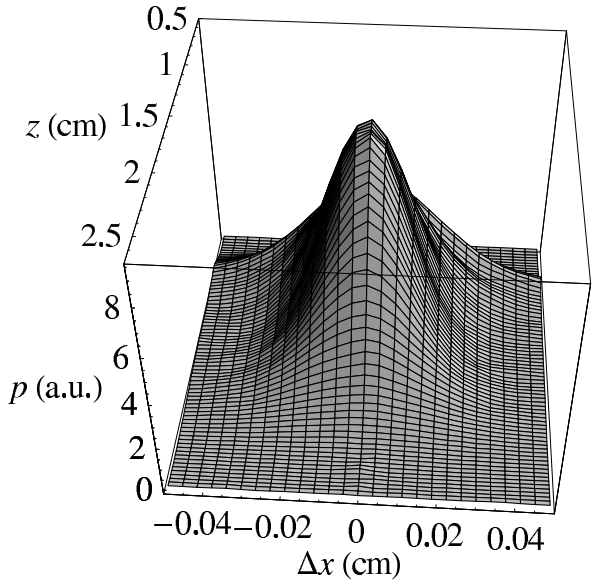}
\caption{\label{deltax}
	Distribution $p(\Delta x,z)$ of horizontal displacements $\Delta x$ and heights $z$ of grains averaged over the vibration period, for displacement measurement time $\Delta t = 1.35$~ms.
	The vertical axis is arbitrary.
}\end{figure}

	The distribution of horizontal displacements $p(\Delta x, z)$ is shown in Fig.~\ref{deltax}, for $\Delta t=1.35$~ms well within the ballistic regime.
	Thus $p(\Delta x/\Delta t, z)$ should approximate the distribution of horizontal velocities $p(v_x,z)$\cite{wildman99}.
	It can be seen that $p(\Delta x,z)$ is symmetric in $\Delta x$ and roughly  gaussian at each height $z$, at least well away from the sample bottom.
	This suggests that the motion is dominantly statistical with minimal large-scale convection.
	The distribution of \emph{vertical} displacements (not shown here) is nearly symmetric at the sample top, but becomes skewed as the sample bottom is approached.
	This reflects the asymmetry between upward moving grains that have recently impacted the cell bottom, and downward moving grains that have not done so.

	We fit the distribution of horizontal displacements $p(\Delta x, z)$ to the form $\exp-[(\Delta x/s_x)^2/2]^\beta$ with $s_x$ and $\beta$ as fitting parameters ($\beta=1$ gives a gaussian).
	The best fit $\beta$ varies from 0.9 at the sample bottom to nearly 1.0 at the sample top, although the deviation of $\beta$ from unity is only marginally significant even at the sample bottom.
	Therefore the deviation of the velocity distribution from gaussian in this three-dimensional system is less than seen experimentally in two dimensions, for which one recent study found $\beta \approx 0.75$\cite{rouyer00}.
	The RMS grain displacements computed from the fits agree to within 10\% with $\Delta x_{rms}$, $\Delta z_{rms}$ computed directly as described above.

\begin{figure}
\includegraphics[width=0.9 \linewidth]{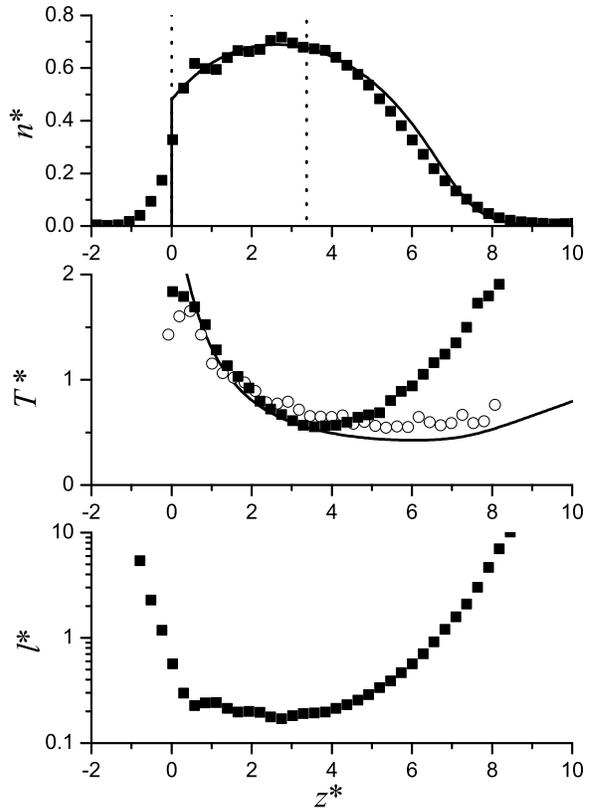}
\caption{\label{tilde}
	(Top) Measured dimensionless number density $n^*$ (filled squares).
	(Center) Measured dimensionless horizontal granular temperature $T_x^*$ (filled squares) and vertical granular temperature $T_z^*$ (open circles).
	(Bottom) Dimensionless mean free path $\ell^*$, computed from the density data.
	These quantities are plotted versus vertical position measured in grain diameters, $z^* = z/\sigma$.
	The smooth curves show the number density $n^*$ and granular temperature $T^*$ calculated from the theory of Ref.~[4], with the  heat current at sample bottom $q_z^*(0)=3.4$ and restitution coefficient $\alpha=0.90$ adjusted to fit the data.
	The vertical dotted lines in the top graph show the extent of the granular sample in the absence of vibration.
}\end{figure}

	To compare the data with theory, the density was normalized using $\int n(z) dz = n_A$ where $n_A=119.6$~cm$^{-2}$ is the areal density\cite{na}.
	The granular temperatures for horizontal and vertical motion were computed from the ballistic-regime ($\Delta t = 1.35$~ms) mean-square displacements as $T_x = m\langle \Delta x^2 \rangle/ \Delta t^2$, $T_z = m\langle \Delta z^2 \rangle/ \Delta t^2$.
	Dimensionless height, number density and temperature variables were computed from $z^* = z/\sigma$, $n^* = n \sigma^3$ and $T_{x,z}^* = T_{x,z}/mg\sigma$ using the values of $\sigma$, $m$ and $g$ given above.
	Figure \ref{tilde} shows the measured $n^*(z^*)$ and $T_{x,z}^*(z^*)$ as well as the dimensionless mean free path $\ell^*(z^*) = \ell/\sigma$, estimated in a manner similar to Ref.~\onlinecite{grossman97}.

	The maximum observed density $n^* \approx 0.7$ should be compared with the values $n^*=1.2,1.41$ for poured random and crystalline close packed monodisperse spheres, respectively\cite{dullien92}.
	In the lower part of the sample ($z^* \leq 4$), the horizontal and vertical granular temperatures are nearly equal to one another and decrease monotonically with height.
	Although energy is supplied to the vertical degrees of freedom by collisions with the cell bottom, the mean free path is much less than the particle diameter ($\ell^* \ll 1$) in this dense portion of the sample, resulting in effective equilibration between $T_x^*$ and $T_z^*$.
	Above $z^*=5$ the vertical temperature $T_z^*$ continues to decrease slowly with height while the horizontal temperature $T_x^*$ increases rapidly, falling far out of equilibrium with $T_z^*$.
	It is reasonable that $T_x^*$ and $T_z^*$ should become unequal where the mean free path is long, but we have not yet identified the physical mechanism that causes $T_x^*$ to increase rapidly with height as measured.

	The smooth curves in Fig.~\ref{tilde} show a fit to the data of the hydrodyamic theory for inelastic hard spheres of Garz\'{o} and Dufty\cite{garzo99}, which is applicable for wider ranges of density and inelasticity than other available theories.
	In inelastic fluids the heat current is ${\bf q} = -\kappa{\bf \nabla}T -\mu{\bf \nabla}n$ where $\kappa$ is the thermal conductivity and $\mu$ is a new transport coefficient that vanishes for elastic systems\cite{soto99,brey01}.
	We have numerically integrated the expressions for $\kappa$, $\mu$, and the pressure $p$ from Ref.~\onlinecite{garzo99} with respect to $z^*$ to obtain the curves shown in Fig.~\ref{tilde}.
	Due to the open top boundary the dimensionless pressure at the bottom is $p^*(0) = (\sigma^2/mg)p(0) = n_A^*$.
	Here $n_A^* = \sigma^2 n_A = 4.04$ is the dimensionless areal number density.
	The upward heat current at the cell bottom $q_z^*(0) = (\sigma^{3/2}/g^{3/2}m)q_z(0)$ and the restitution coefficient $\alpha$ are used as fitting parameters.
	Presumably $q_z^*(0)$ is set by the amplitude and frequency of cell vibrations.
	The density at the bottom $n^*(0)$ is set to give vanishing heat current at the cell top ($q_z^*(z^*) \rightarrow 0$ as $z^* \rightarrow +\infty$).

	Once $q_z^*(0)$, $\alpha$ and the sample bottom location are adjusted to fit the experimental density profile $n^*(z^*)$, the granular temperature profile $T^*(z^*)$ is predicted by the theory with no further adjustable parameters.
	As shown in Fig.~\ref{tilde}, the \emph{vertical} temperature $T_z^*$ agrees well with the theory over the entire measured range of $z^*$.
	However, the \emph{horizontal} temperature $T_x^*$ deviates from the fit for $z^*>5$.
	We find no choice of fitting parameters that duplicates the rapid increase of $T_x^*$ seen experimentally.
	Therefore our data do not prove $\mu \ne 0$, which is what allows the theoretical profile to have $\partial T^*/\partial z^* >0$.

	Are the fitted parameter values $q_z^*(0)=3.4$, $\alpha=0.90$ physically reasonable?
	For an open vibrated system the power input is $Q=2Nmg\langle V\rangle$ where $Nmg$ is the total sample weight and $\langle V\rangle$ is the average upward cell velocity during grain impacts\cite{macnamara97}.
	Combined with the fitted $q_z^*(0)$ this gives $\langle V\rangle = 5.65$~cm/s $=0.12v_0$, where $v_0$ is the peak cell speed.
	This value for $\langle V\rangle$ appears reasonable as most of the collisions with the cell bottom occur near the top of the vibration stroke (Fig.~\ref{density}).

	We have attempted to measure $\alpha$ for the grains used in this study by videotaping their trajectories when dropped onto a hard surface.
	The apparent $\alpha$ varies from trial to trial, as the irregular shape causes conversion of some translational energy into rotational energy at impact.
	For an impact velocity of 44~cm/s (larger than is typical in the fluidized sample) the mean apparent $\alpha$ measured in this way was 0.79 and the largest value measured across all trials was 0.92.
	While not an accurate measurement of the restitution coefficient in the fluidized state, these results are at least consistent with the fitted value.

	In summary, we have used NMR methods to measure three-dimensional grain motion and density and granular temperature profiles for a vibrofluidized granular system composed of irregularly shaped mustard seeds.
	The dense lower region of the sample is well described by a hydrodynamic theory for inelastic hard spheres.
	Thus, the continuum theory appears accurate in spite of the small number of grains in the sample and the steep granular temperature gradient.
	In the upper region of the sample the mean free path becomes long, and the horizontal and vertical grain velocities become decoupled in a manner that is not described by the theory.

	We thank N. Menon for useful discussions.  This work was supported by NSF Grant No. CTS 9980194.


\begin{references}
\bibitem{alpha}The restitution coefficient $\alpha \in [0,1]$ is the ratio of final to initial speeds for two colliding grains in their center of mass frame.
\bibitem{vac}The cell evacuation reduces the typical grain Reynolds number from 14 to $4\times 10^{-3}$.  The calculated energy losses due to air drag are much less than collision losses for the pressure used.
\bibitem{na}To correct for excluded volume at the cell wall $n_A$ was deduced by measuring the height of the sample without vibration, assumed to have the number density for a poured random packing $n\sigma^3 \approx 1.20$\cite{dullien92}.
\end{references}
\end{document}